\documentclass[sigconf,10pt]{acmart}

\usepackage{makecell}

\acmYear{2021}
\copyrightyear{2021}

\begin{document}
	
	\title{BuffetFS: Serve Yourself Permission Checks without Remote Procedure Calls}
	
	\author{Yanliang Zou}
	\affiliation{%
		\institution{ShanghaiTech University}
		\city{Shanghai}
		\state{China}
	}
	\email{zouyl@shanghaitech.edu.cn}

	\author{Bin Yang}
	\affiliation{%
		\institution{Shandong University}
		\city{Jinan}
		\state{China}
	}
	\email{bin.yang@mail.sdu.edu.cn}
	
	\author{Jian Zhang}
	\affiliation{%
		\institution{ShanghaiTech Unviersity}
		\city{Shanghai}
		\state{China}
	}
	\email{zhangjian@shanghaitech.edu.cn}
	
	\author{Wei Xue}
	\affiliation{%
		\institution{Tsinghua University}
		\city{Beijing}
		\state{China}
	}
	\email{xuewei@tsinghua.edu.cn}
	
	\author{Shu Yin}
	\affiliation{%
		\institution{ShanghaiTech Unviersity}
		\city{Shanghai}
		\state{China}
	}
	\email{yinshu@shanghaitech.edu.cn}

	\renewcommand{\shortauthors}{Y. Zou et al.}
	
	\begin{abstract}
       The remote procedure call  (a.k.a. RPC) latency becomes increasingly significant in a distributed file system. We propose BuffetFS, a user-level file system that optimizes I/O performance by eliminating the RPCs caused by \texttt{open()} operation. By leveraging \texttt{open()} from file servers to clients, BuffetFS can restrain the procedure calls for permission checks locally, hence avoid RPCs during the initial stage to access a file. BuffetFS can further reduce response time when users are accessing a large number of small files. We implement a BuffetFS prototype and integrate it into a storage cluster. Our preliminary evaluation results show that BuffetFS can offer up to 70\% performance gain compared to the Lustre file system. 
		
	\end{abstract}
	
	
	%
	%

	
	\maketitle
	
	\sloppy
	
	\section{Introduction}

    Modern distributed file systems usually decouple metadata access from read and write to exaggerate the throughput of data storage nodes and hide the latency of metadata operations in data transfers. This mechanism works well when multiple users attempt to read or write large files from distributed file systems as the metadata-to-data ratio is low. The time cost of metadata access is small enough to be neglected when files are large, but when the files are small, the remote procedure call (RPC) latency of accessing metadata becomes significant. As the numbers of small files are estimated to soon achieve and exceed billions, how to optimize RPC latency to metadata is a known challenge for existing distributed file systems. While distributed file system metadata servers generate RPCs for many functions, our focus in this paper is on the access permission operation of the file system during \texttt{open()}.
	
    Distributed file systems usually perform verification of access control on a given pathname argument before storage servers can provide proper file operations. In other words, access to the file specified by a pathname requires permission to check its ownership and its grouping information and permission to access the data in a mixed mode of \texttt{read}, \texttt{write}, and \texttt{execute}. We call this file access control the permission check. The permission check is traditionally the first operation for all file system operations on files and is the basis of most user's understanding of file system access controls. Modern distributed file systems usually apply centralized single-node or multi-node metadata services to break the bottleneck from the serialization of permission checks. While distributed file systems such as Lustre\cite{lustre}, GPFS\cite{GPFS}, GlusterFS\cite{GlusterFS}, and GFS\cite{GFS} benefit from better scalability and concurrent data access, the independent metadata services introduce additional RPCs to access a file. For example, Lustre usually generates at least three round-trip RPCs to access a file: \texttt{open()}, \texttt{read()} or \texttt{write()}, and \texttt{close()}. Although the RPC by \texttt{close()} can be executed asynchronously, RPCs by \texttt{open()} and \texttt{read()}/\texttt{write()} still cost a constant latency to Lustre.
    

	
    Current studies\cite{vNFS,Chen-SIGMETRICS2015,Chen-SISTOR2016} usually try to reduce the number of RPCs by combining \texttt{open()} with \texttt{read()}/\texttt{write()} requests or aggregating a number of requests. However, these attempts require additional efforts to modify applications' API due to their incapability with the standard POSIX API\cite{Bok2017AnED}. Some studies aim to optimize the metadata management efficiency by flattening metadata via a key-value store or replicating metadata across WAN servers\cite{calvinfs}\cite{flatten_meta}\cite{ShardFS2015}. Although these solutions mitigate the overhead by hierarchical metadata operations, they barely reduce the number of RPCs by metadata access, which is mainly causes existing distributed file systems to perform poorly to access small files. Studies also make efforts to improve the efficiency of assessing small files by diminishing RPCs. For example, the Lustre group proposes Data on MDT (a.k.a. DoM) mechanism that aims at reducing RPCs for data accessing\cite{DoM}. DoM optimizes the number of RPCs to read small files by storing them on the metadata servers. However, when the number of small files keeps growing, this approach may drain precious storage capacity on the metadata servers and aggravate the already serious bottleneck of metadata access. Besides, this approach is not write-friendly because all the writes to small files will congest the metadata servers. 
    
    \texttt{open()} is the left-aside operation that generates an RPC for every file access. \texttt{open()} usually consists of two steps-- Step 1: checks the permission of the requested file and Step 2: records the open status of the file. Upon every \texttt{open()}, a client node (hereafter referred to client) in a distributed file system issues one RPC to check the permission on the server while the server performs both the two steps and returns the information to the client. We can foresee that avoiding the RPC that checks files permission between the client and metadata server could improve the latency of distributed file systems for accessing enormous small files. Yet, it is very challenging in existing distributed file systems to diminish such RPC since \texttt{open()} has to access the metadata server via the network to initial operations to a file. 
	
    We propose BuffetFS, a user-level distributed file system that conceals the RPC by \texttt{open()} during the file accessing initial. BuffetFS dis-aggregates \texttt{open()} then leverages the permission check (Step 1) to the client-side for better responsiveness and defers the file offset and flags recording (Step 2) to its successor operations that need to contact metadata servers. In this case, BuffetFS restrain the number of RPCs to one for actual data accesses (i.e. \texttt{read()} or \texttt{write()}), which alleviates the latency for accessing a large number of small files. The core part of BuffetFS is a mechanism to attach files' permission information to their parent directory. Besides inode numbers and name strings, the BeffetFS directory also contains the permission information of all the files and sub-directories that belong to it. Each client in BuffetFS maintains an incomplete directory tree structure that consists of directories accessed before and their children. Besides, each client holds the complete permission information in the directory tree. By doing so, BuffetFS balances the response time for \texttt{open()} and the storage capacity. As a price, the servers must keep all related clients updated when applications modify the permission of a file/directory, which usually doesn't occur frequently. 
	
	
    As the \texttt{open()} is dis-aggregated, BuffetFS only needs to manage servers that store files and directories data and does not require a centralized metadata server. Hence a decentralized distributed file system becomes possible via BuffetFS. The clients can locate a file or a directory from its inode number that consists of a host-ID and a unique file-ID. Thus, a client can check files' permission by itself and access the files without requesting their location and metadata from other clients. 
	
    The main contributions of this paper include: (1) we propose BuffetFS, a user-level file system to optimize I/O performance to enormous small files in a distributed environment. BuffetFS is designed to reduce \texttt{open()} latency by eliminating RPCs for permission check;  (2) we implement a BuffetFS prototype and a decentralized distributed file system sandbox; (3) a comprehensive experimental study is provided to evaluate the efficacy of the BuffetFS prototype.
	
	The rest of the paper is organized as follows. Section~\ref{section:Background} provides the background and motivation of this research. The design and implementation details of BuffetFS are presented in Section~\ref{section:Design}, which is followed by an evaluation of BuffetFS shown in Section~\ref{section:preliminary}. 
	Section~\ref{section:relatedwork} summarizes the related work. Finally, Section~\ref{section:conclusion} concludes this paper.
	
	
	\section{Motivation and Background}
	\label{section:Background}
	\subsection{Motivation}
	Besides providing I/O services for traditional compute-intensive applications such as scientific simulations, the modern distributed file systems start to support applications like machine learning and Big Data analysis, which access enormous small files. 
	According to our observation on an object storage server (OSS) of a Lustre cluster in TaihuLight supercomputing center, more than 90\% RPCs come from accessing small files. This Lustre cluster is mainly utilized by some machine learning studies and the Beacon system\cite{beacon}, which is an I/O surveillance system of the whole TaihuLight supercomputer. The accessing statistic of this Lustre cluster indicates that more than 70\% of metadata operations are \texttt{open()} and \texttt{close()}. As Lustre can execute \texttt{close()} in an asynchronous way, \texttt{open()} becomes the major operation that causes remote I/O latency for accessing small files. 
	
	
	
	We further notice that metadata of directories can be cached on the client-side to optimize the response time, however, the permission of the target file in \texttt{open()} can not be cached unless the file is accessed before. How to optimize RPC latency by permission check motivates us to design a scheme to leverage this operation to the client side. 
	
	\subsection{Background of \texttt{open()} operation}
	In this section, we briefly explain the general \texttt{open()} operation in two dominant types of file systems -- local file systems and network-connected file systems. 
	
	For local file systems, the \texttt{open()} function invoked by any applications will be interpreted as an \texttt{open()} system call to the kernel. The kernel then parses the path string and finds dentries and inode objects for each involved path component in turn. For every step of directory component traversal, the kernel has to check the permission of the current component before proceeding to the next step. Only upon the granted permission of the current component will allow the kernel to move on to the next step. 
	Upon reaching the target component, the kernel checks its complete permission according to the \texttt{open()} flags while for its parent directory components, the kernel checks the execution permission only. The kernel then marks the referenced file as opened and returns a unique file descriptor to the application. Other file objects such as the inode object, file object, and superblock object will be updated consequently. 
	
	As for network-connected file systems such as distributed file systems, the kernel reserves APIs of the steps mentioned in the local file systems (i.e., request for metadata, permission checks). For example, a Lustre client sets a flag for each dentry to mark its availability to deal with concurrent modification to the same directory or files by other remote Lustre clients. Distributed file systems usually maintain a global lock manager to preserve the data and metadata integrity of files. For example, Lustre implements a  Lustre Distributed Lock Manager (a.k.a LDLM) to ensure the data consistency\cite{lustre-internal}). One side-effect of global lock management is that it introduces external permission management. Hence distributed file systems maintain metadata and a list of open files on the server-side, which requires one RPC to perform \texttt{open()} on every objective file.
	
	However, there is no need to perform all the steps of \texttt{open()} on the server-side. Recall the two major steps in \texttt{open()} (i.e., check permissions and record file status), only the permission check has to executed immediately while the status record can be postponed and be performed asynchronously. If a client can perform a permission check for \texttt{open()}, one RPC to the server can be concealed.
	
	\section{The Design of BuffetFS}
	\label{section:Design}
	\subsection{BuffetFS Architecture}
\begin{figure}[htpb]
		\centering
		\includegraphics[width=0.98\columnwidth]{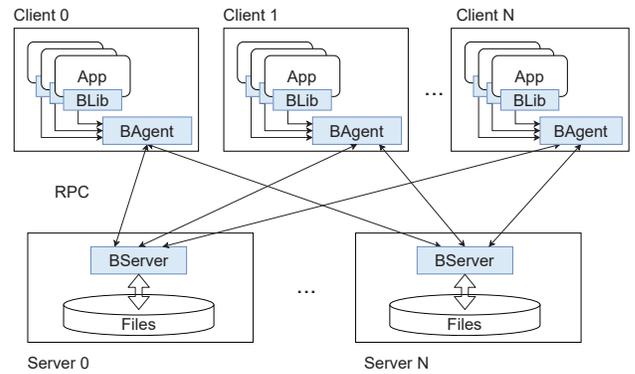}
		\caption{BuffetFS Architecture}
		\label{figure:architecture}
	\end{figure}

	Fig. \ref{figure:architecture} illustrates the architecture of BuffetFS, Serving as an userspace distributed file system, BuffetFS sits between the applications and an underlying file system. BuffertFS consists of three components: a BuffetFS Library (\textbf{BLib}), a BuffetFS Agent (\textbf{BAgent}), and a BuffetFS Server (\textbf{BServer}).
	
	BLib serves as a dynamic library that intercepts and redirects the POSIX I/O requests from applications to BAgent. Locates in a client, a BAgent maintains an incomplete directory tree where each tree node occupies the permission status and the pointers of files. A BAgent also maintains a corresponding context to a user process including the PID, file descriptors, and file objects. Every client can only have one BAgent.
	
	BServer is deployed on a server node to manage actual file data. It collects file access requests from all the clients and maintains the files' metadata. For the \texttt{open()} operation, a BServer maintains a list of opened files to ensure data consistency for concurrent file modifications from multiple clients.
	
	\subsection{Namespace and Metadata Handling}
    BuffetFS does not have a centralized metadata server since the permission checks are leveraged to clients. To manage a global namespace, BuffetFS re-modifies the inode to contain three segments: (1) a \textit{hostID}, representing the server that stores the actual file data; (2) a \textit{fileID}, which is a unique number to identify a file on the corresponding BServer locally;  and (3) a \textit{version number} of the server, which is designed to record exceptions of a server (e.g., reboot or restore). The BAgent on each client maintains a local configuration file that maps a tuple (a hostID and a version number) to a server address. Thus, every single inode number on a client can identify the location of a corresponding file on a server.
	
	Each file in BuffetFS owns a pair of front-end/ back-end metadata, where the front-end metadata is related to the client user and the back-end metadata on the server-side is for the management of actual files. Some front-end metadata will be stored in the extended attributes of the actual file in BServer to handle BuffetFS inode numbers and clients' permission. Other than that, both the front-end and back-end metadata store the same information including last access time, modify time and create time. 
	
	
	In addition to the regular inode number and name strings for files and sub-directory, BuffetFS uses ten extra bytes for each directory entry to store the permission information. The total extra bytes for a complete directory is commonly no more than hundreds of bytes, which shouldn't be a problem. 
	
	\subsection{Buffet I/O control flow}
	In this subsection, we discuss the control flow details of BuffetFS.
	
\begin{figure}[htpb]
		\centering
		\includegraphics[width=0.98\columnwidth]{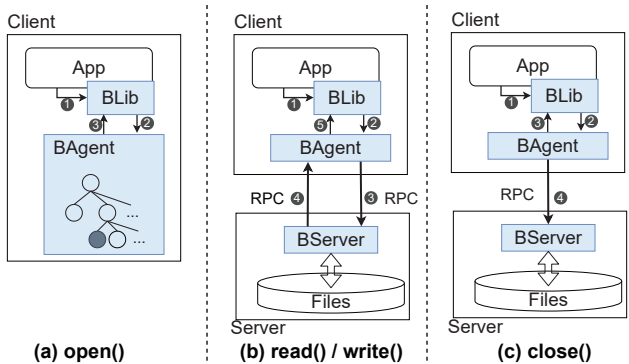}
		\caption{BuffetFS control flow
		}
		\label{figure:reading_flow}
	\end{figure}

	Presented in Fig.~\ref{figure:reading_flow}(a) Upon an application issues \texttt{open()}, BuffetFS first intercepts the operation to a BLib (a-1), the BLib then redirects the \texttt{open()} to a BAgent (a-2) and monitors the returned file descriptor (fd) from the BAgent (a-3). The BAgent traverses the path string in the local cached directory tree to find the corresponding tree node. If the tree node is cached, the BAgent obtains the permission locally; otherwise, it obtains the complete location and permission data of the parent node then extends the cached directory tree. As for the referenced file, the BAgent does not need to generate one RPC to collect its metadata from the server since the file's permission is recorded in its parent directory. For example, when a user tries to open the file \texttt{foo} with the pathname \texttt{``/a/b/foo"} while a BAgent has cached the directories \texttt{a/} and \texttt{b/} locally in advance, the BAgent first obtains the data of \texttt{b/} and inserts all the \texttt{b/}'s children to the cached tree to replenish \texttt{foo} information, then proceeds the permission check of \texttt{foo}. In the end, the BAgent picks and returns a valid fd to BLib. Meanwhile, the BAgent marks the file object as incomplete-opened until the BServer finishes the rest of \texttt{open()} operations.   
	
	Once the BLib receives and returns the fd the application, BuffetBF can proceed to provide \texttt{read()} or \texttt{write()} operation (demonstrated in Fig.~\ref{figure:reading_flow}(b)). A BAgent attaches the incomplete-opened flag to the first read or write request cauesd by the same process (b-2) and sends them to the corresponding BServer based on the BuffetFS inode number (b-3). After parsing the RPCs, BServer executes the rest operations of \texttt{open()} operations (i.e. updates the opened-file list) then the read request. It identifies the back-end file with the fileID and returns the file data to the BAgent via one RPC (b-4). The BAgent finally returns the requested data (for \texttt{read()}) or the statue (for \texttt{write()}) to the application (b-5).
	
    As for \texttt{close()}, the BAgent returns a signal immediately and performs an RPC asynchronously to inform the corresponding the BServer to wrap up operations by removing the file object from the opened-file list.
	
    To sum up, only \texttt{read()} or \texttt{write()} RPCs impacts the whole latency while the RPC by \texttt{close()} can be hided asynchronously.
	
	\subsection{Metadata Modification and Consistency}
	Since BuffetFS executes permission checks on the client-side, it introduces overhead for modifying file permission. Upon any permission updates, a BServer has to inform all the related clients to invalid the corresponding cache entries then execute the permission changes.
	
	For each directory, a BServer records a list of clients that cache the directory data. Upon any file permission changes, a BServer has the big picture of all the related clients. Therefore, the BServer produces RPCs to inform the corresponding BAgents to invalid the involved tree nodes. After receiving all the responses from clients, the BServer then executes the permission modification. 
	
	If a BAgent tries to access an invalid tree node, it will ask for the updated permission from the BServer. This mechanism ensures the strong consistency of the metadata. 
	Other metadata modifications, such as changing file name and file migration, may cause similar overheads among BuffetFS and DFS theoretically. They all need to ask the related clients to invalidate corresponding local cached metadata.
	
	
	
	

    \section{BuffetFS Evaluation}
    \label{section:preliminary}
    We carry out our evaluation on a Sunway TaihuLight HPC testbed cluster to test the access latency and concurrency. Each node is equipped with two 2.6GHz 8-core 16-thread Intel Xeon CPU and 64GB RAM, running CentOS v7.5. The cluster is built on a Lustre (v2.10) file system with 1 MDS and 4 OSSes interconnecting with InfiniBand. Lustre's storage targets consist of 12 HDDs which are arranged as two groups of RAID6 by a Sugon DS800-F20 storage system providing 71TB HDD storage capacity. We deploy BuffetFS laying over ext4 in the cluster and compare its performance with Lustre.
	
	We arrange our tests with three groups: BuffetFS, Lustre-Normal, and Lustre-DoM. While Lustre-Normal stands for the common utilization of Lustre, particularly, Lustre-DoM presents the performance of tests running on Lustre with DoM mode (Data on MDS). With DoM, a small file's data can be stored on MDS and clients can get both metadata and data on MDS (please see Section~\ref{section:relatedwork}). 
	
\begin{figure}[htpb]
		\centering
		\includegraphics[width=0.98\columnwidth]{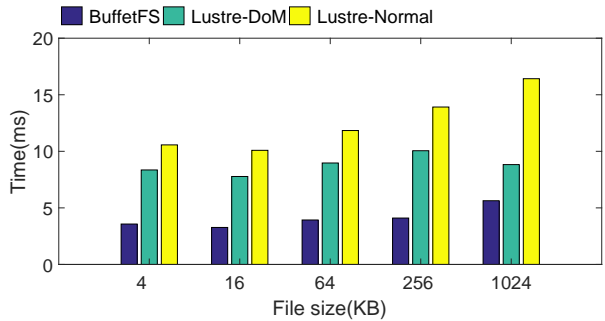}
		\caption{Latency of accessing files (single process)}.
		\label{figure:single_file}
	\end{figure}

	Figure~\ref{figure:single_file} shows the performance of accessing a single small file including three operations: \texttt{open()}, \texttt{read()}, and \texttt{close()}. BuffetFS performs the lowest latency in the test compared to the two Lustre cases. Firstly it conceals one round RPCs caused by \texttt{open()} so that only the \texttt{read()} RPCs affect the latency (\texttt{close()} will be executed asynchronously). Another reason is that BuffetFS arranges files locks inside the BServer for concurrency while Lustre arranges its distributed file locks among all of its clients.  
	
\begin{figure}[htpb]
		\centering
		\includegraphics[width=0.98\columnwidth]{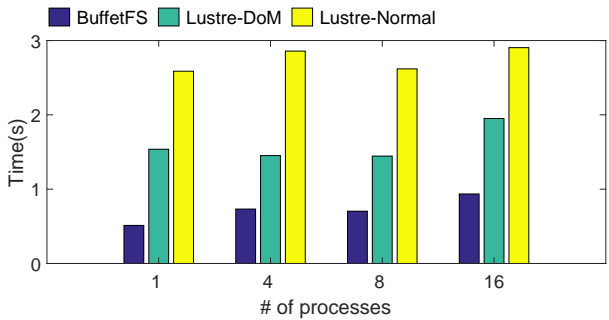}
		\caption{Total execution time of concurrent accessing multiple files (1,000 files per process, file size: 4KB, file quantity: 100,000)}
		\label{figure:multi_files}
	\end{figure}

	Further, we explore the performance of the three groups on concurrent access (Figure~\ref{figure:multi_files}). We fork different numbers of processes each of which randomly accesses 1000 files among 100000 4KB files. To eliminate the effect of data cache and other internal mechanisms in Lustre, we regenerate the files set for each test.
	
	BuffetFS requests for the directory data once and built the directory tree on the client, which means that the following access to other files in the same directory can benefit from the cached directory tree without applying metadata from BServer. Differently, the two Lustre cases have to request metadata from MDS for each \texttt{open()} operation. And BuffetFS presents a standout performance in this case.

    
    \section{Related Work}
    \label{section:relatedwork}
	In this section, we discuss other studies focusing on optimizing metadata access to improve the performance of distributed file systems. 
	
	IndexFS\cite{IndexFS} is a layered cluster file system to optimize metadata service for common distributed file systems (e.g., HDFS\cite{Hadoop}, Lustre\cite{lustre}). It caches partition directory tree on the clients and uses a short-term lease for each cached directory entry to ensure its consistency. What the client cache is interested in are directories entries visited before. In other words, when the client accesses a file, components on the file path will be cached on the client. However, this strategy does not always effective, for example, Lustre itself will keep directory entry valid on the client after access. Besides, the open operations can not benefit from the cache of IndexFS since it does not cache the last component. 
	
	As we mentioned above, Lustre\cite{Lustre-src} keeps directory entries valid on a client after accessed. The following visits to the valid entries do not need to contact the Metadata Server (MDS). Besides, the Data on MDS (DoM)\cite{DoM} mechanism speedups small file accessing for Lustre. It stores file data on MDS as an extended attribute of the file. With DoM, when an \texttt{open()} request comes, MDS will attach the file data to the returned RPC so that the client does not need to access the Object Storage Server (OSS) anymore. But DoM only optimize the \texttt{open()-read()-close()} operations while \texttt{open()-write()-close()} does not benefit from it. Further, DoM occupies the space of MDS which is expensive to Lustre.
	
	
	\section{Conclusions}
    \label{section:conclusion}
	In this paper, we identified a challenge of data access faced by many distributed file systems for accessing a large number of small files. To solve it, we developed a prototype of a user-level file system called BuffetFS which is then integrated into a cluster. Further, we applied BufferFS to optimize RPCs for permission check in \texttt{open()}. Our experimental results demonstrate that BufferFS can noticeably reduce the latency of \texttt{open()} in distributed file systems. 
	
	BuffetFS currently can support major I/O operations including \texttt{open()}, \texttt{read()}, \texttt{write()}, and \texttt{close}. As an ongoing work, we are promoting BuffetFS to support POSIX I/O API and internal optimizations such as full metadata caching.
	
	
	\bibliographystyle{ACM-Reference-Format}
	\bibliography{myref}
\end{document}